\documentclass[namedreferences]{solarphysics}

\usepackage[hyperref,optionalrh,showbiblabels]{spr-sola-addons} 
\usepackage[optionalrh]{spr-sola-addons} 
\usepackage{graphicx}        
\usepackage{breakurl}        
\usepackage{solaheader}	
\newcommand{\solareceiveddate}{24 February 2018} 
\newcommand{\solaaccepteddate}{20 July 2018}
\begin{document}
\begin{article}
\begin{opening}
\title{Tilt of sunspot bipoles in Solar Cycles 15 to 24}

\author[addressref={aff1},email={k.tlatova@mail.ru}]{\inits{K. A.}\fnm{Ksenia}~\lnm{Tlatova}\orcid{0000-0001-8345-6801}}
\author[addressref={aff2,aff3},email={tlatov@mail.ru}]{\inits{A.G.}\fnm{Andrey}~\lnm{ Tlatov}\orcid{0000-0003-1545-2125}}
\author[addressref={aff4,aff5},corref,email={apevtsov@nso.edu}]{\inits{A.A.}\fnm{Alexei}~\lnm{Pevtsov}\orcid{0000-0003-0489-0920}}
\author[addressref={aff5},email={kalevi.mursula@oulu.fi}]{\inits{K.}\fnm{Kalevi}~\lnm{Mursula}\orcid{0000-0003-4892-5056}}
\author[addressref={aff2},email={}]{\inits{V.V.}\fnm{Valeria}~\lnm{Vasil'eva}}
\author[addressref={aff5},email={Elina.Heikkinen@oulu.fi}]{\inits{E.}\fnm{Elina}~\lnm{Heikkinen}}
\author[addressref={aff4},email={lbertello@nso.edu}]{\inits{L.}\fnm{Luca}~\lnm{Bertello}}
\author[addressref={aff4,aff5},email={apevtsov@nmsu.edu}]{\inits{A.A.}\fnm{Alexander}~\lnm{Pevtsov}}
\author[addressref={aff5},email={Ilpo.Virtanen@oulu.fi}]{\inits{I.I.}\fnm{Ilpo}~\lnm{Virtanen}\orcid{0000-0001-7258-4453}}
\author[addressref={aff6},email={ninakarachik@mail.ru}]{\inits{N.V.}\fnm{Nina}~\lnm{Karachik}}

\address[id=aff1]{St. Petersburg State University, Saint Petersburg, Russian Federation}
\address[id=aff2]
{Kislovodsk Mountain Astronomical Station of Pulkovo Observatory,
Kislovodsk, Russian Federation}
\address[id=aff3]
{Kalmyk State University, Elista, 358009, Russian Federation}
\address[id=aff4]
{National Solar Observatory, 
3665 Discovery Drive, 3rd Floor,
Boulder, CO 80303 USA}
\address[id=aff5]
{ReSoLVE Centre of Excellence, Space Climate research unit, University of Oulu, POB 3000, FIN-90014, Oulu, Finland
}
\address[id=aff6]{
Astronomical Institute AS RUz, 33 Astronomicheskaya ul., Tashkent, 100052, Uzbekistan}
\begin{abstract}

We use recently digitized sunspot drawings from Mount Wilson 
Observatory to investigate the latitudinal dependence of tilt angles 
of active regions and its change with solar cycle. The drawings cover
the period from 1917 to present and contain information about polarity 
and strength of magnetic field in sunspots. We identify clusters of 
sunspots of same polarity, and used these clusters to form
``bipole pairs''. The orientation of these bipole pairs was used to
measure their tilts. We find that the 
latitudinal profile of tilts does not monotonically increase with 
latitude as most previous studies assumed, but instead, it shows a clear 
maximum at about 25--30 degree latitudes. 
Functional dependence of tilt ($\gamma$) on latitude ($\varphi$) was found to be 
$\gamma = (0.20\pm 0.08) \sin (2.80 \varphi) + (-0.00\pm 0.06)$.
We also find that latitudinal dependence 
of tilts varies from one solar cycle to another, but larger tilts
do not seem to result in stronger solar cycles. Finally, we find
the presence of a systematic offset in
tilt of active regions (non-zero tilts at the equator), with odd cycles exhibiting negative offset
and even cycles showing the positive offset.

\end{abstract}

\end{opening}

\section{Introduction} \label{sec:intro}

Orientation of solar bipoles, as defined by a line connecting their leading 
and following parts, exhibits a slight systematic tilt relative to the 
direction of solar circles of constant latitude (parallels) with trailing part being 
situated at higher latitudes as compared with their leading part. Tilt 
angles show a tendency to increase with the latitude. This pattern, dubbed 
``Joy's law'' by H. Zirin (\citeyear{Zirin1988}), was first described by 
\inlinecite{Hale.etal1919}. In  the framework of surface flux transport models, the 
active region tilt is one of the important ingredients that affects 
the strength of polar field and therefore, is important for the amplitude of 
the next solar 
magnetic cycle (\textit{e.g.}, \opencite{Cameron.etal2010}). Joy's law was extensively 
studied using both white light and magnetic field observations 
(\textit{e.g.}, \opencite{Howard1991,Sivaraman.etal1999,McClintock.etal2014,McClintock.Norton2016}, for review, see \opencite{Pevtsov.etal2014}). 
Current 
interpretations of active regions' tilt include three underlining 
causes: toroidal field orientation (\textit{e.g.}, \opencite{Babcock1961,Norton.Gilman2005}), action of the Coriolis force 
(\textit{e.g.}, \opencite{Fisher.etal2000}), and signature of kink-instability 
(\textit{e.g.}, \opencite{Leighton1969,Longcope.etal1999,Holder.etal2004}). 

Unlike the model
predictions, the observational evidence about the importance of the tilt angles for determining the
strength of the next solar cycle is inconclusive. Several authors (\textit{e.g.}, 
\opencite{Sivaraman.etal1999,Dasi-Espuig.etal2010}) reported that for Cycles 
16--21 the average tilt angles, normalized by latitude, correlate with 
the amplitude of the next solar cycle. On the other hand, later studies (\textit{e.g.}, \opencite{Ivanov2012,Mcclintock.Norton2013}) could not reproduce the results of  
\inlinecite{Dasi-Espuig.etal2010}, and later, \inlinecite{Dasi-Espuig.etal2013} 
revised their earlier results.  \inlinecite{Mcclintock.Norton2013} indicated 
that cycle variation of tilt angles may differ for the two solar 
hemispheres. Indeed, \inlinecite{Li.Ulrich2012} found the least-square linear 
fit to their measurements of tilts derived from magnetogram data during 
1974--2013 (Cycles 21--24) as 
$\gamma = (0.5 \pm 0.2)\varphi - (0.9^\circ \pm 0.3^\circ)$. They 
concluded that the observed non-zero offset reflects the asymmetry of tilts 
between Southern and Northern hemispheres.
The dependency of tilt ($\gamma$) on latitude ($\varphi$) is typically 
fitted by a function of latitude, $\gamma \propto f(\varphi)$ or sine of 
latitude, 
$\gamma \propto f(\sin \varphi)$. Functional dependency 
is often (but not always, see examples in Table 1 in \opencite{Pevtsov.etal2014}) assumed to 
have a zero tilt angle at the equator, $\gamma \propto f(0) 
\equiv 0$. \citeauthor{Pevtsov.etal2014} (\citeyear{Pevtsov.etal2014}, and references therein) noted that 
both of these assumptions may require 
additional studies: some datasets appear showing a maximum tilt angle in some
latitudes, and when the fit is not constrained by the assumption of zero 
crossing at the equator, it often shows a non-zero offset. The presence of 
such an offset may indicate that our present understanding of tilt of 
active regions is incomplete. 

In this paper, we use the magnetic field 
observations in sunspots from the Mount Wilson Observatory (MWO) to further investigate the latitudinal dependence of tilt angles in Cycles 15--24. 
In Section \ref{sec:data} we 
describe our dataset and the approach in determining the proxy for tilt 
angles. Section \ref{sec:1hemi} discusses variation of tilt with solar 
latitude for both hemispheres. In Section \ref{sec:2hemi} we compare
latitudinal variation of tilt for different solar cycles, and in 
Section \ref{sec:discuss} we discuss our findings.

\section{Data} \label{sec:data}

We use data derived from daily sunspot drawings taken at Mount Wilson 
Observatory (MWO). This MWO program started in early 1917, and it 
continues till present with some short interruptions in more recent 
years due to funding shortages. A typical drawing (for a graphic 
example, see Figure 1 in \opencite{Pevtsov.Clette2017}) contains information about 
location of sunspot, its approximate size (penumbra and umbra) as well as 
polarity and maximum 
field strength of its magnetic field. The drawings were digitized 
(tabulated) using software package developed by us (\textit{e.g.}, \opencite{Tlatova.etal2015}).The digitization included the date and time of 
observations, heliographic coordinates of each umbra, its area, the 
strength, and polarity of its magnetic field. The digitized dataset 
employed in this article contains 20318 days of observations, which cover 
period between January 1917 and October 2016. Total number of features (sunspots 
and pores) on these images is about 5$\times$10$^5$.

The results of the initial digitization (version 2017\_11\_01\_AT\_mwo, 
used here) may contain small number of errors related to identification 
of solar limbs on drawings, incorrect 
time stamps, and in rare cases, incorrect polarity recordings from the 
original drawings. Such errors are not unusual given the size of our 
dataset and various subjective factors related to both observations 
(multiple observers with different hand-writing and using different 
recording techniques) and manual digitization. A detailed study to 
quantify and correct some of these errors is currently underway 
(Pevtsov \textit{et al.} 2018, in preparation), but it is clear that the effect 
of these errors is rather small.

One clear advantage of the MWO sunspot drawings as compared with all 
previous studies of tilt angles using historical data is that  
knowing the polarities of sunspots allows a better determination of sunspot 
pairs forming a group. \inlinecite{Pevtsov.etal2014} provided examples of 
group mis-identification when only white light images are used.

Sunspot groups are identified on the original MWO drawings by the MWO group number 
and heliographic coordinates of the center of the group, but these data were not 
included in the initial digitization used here. Hence, we turned to an 
alternative procedure of identifying bipole pairs of opposite polarity. The detailed discussion of early version of this 
algorithm is presented in \inlinecite{Tlatov.etal2010}. We slightly modified the 
algorithm by adding a step to identify clusters of sunspots of positive and negative polarity. 
To identify a cluster, we start from the largest (by area) 
sunspot of corresponding (say, positive) polarity, and search for same polarity sunspots 
within 10 degrees in longitude and 7 degrees in latitude.  The heliographic coordinates
of center of each cluster are computed as mean location of all sunspots within cluster 
weighted by their areas. The clusters are used to form bipolar pairs. At this step, 
the algorithm starts from the center of an arbitrary cluster of positive polarity and 
searches for a negative polarity cluster, whose center is at the distance within 15 degrees 
in longitude and 7 degrees in latitude. The procedure is 
repeated to search for a closest cluster of positive 
polarity using as starting point cluster of negative polarity identified on previous step.
If the closest positive polarity cluster found on this step coincides with the starting 
positive polarity cluster, both positive and negative polarity clusters are 
marked as a pair, and removed from the list of elements for future 
searches. Similar procedures had been successfully used in several past 
studies (\textit{e.g.}, \opencite{Sattarov.etal2002}). 
Once the bipole pair is identified, the tilt angle of its magnetic axis with respect to the constant latitude circle is calculated.
We adopted a ``classical'' definition of active region tilt as 
angle between the line connecting leading and trailing polarity sunspots of 
a group and the solar equator. This definition corresponds 
to Figure 1b in \inlinecite{Li.Ulrich2012}. In this reference frame, active regions following
the Joy's law will have positive/negative tilt in the northern/southern hemisphere. 
Mean latitude (and longitude) of groups was determined as the mean of heliographic 
coordinates of two clusters comprising the group.

\begin{figure}[ht]
\includegraphics[width = \textwidth]{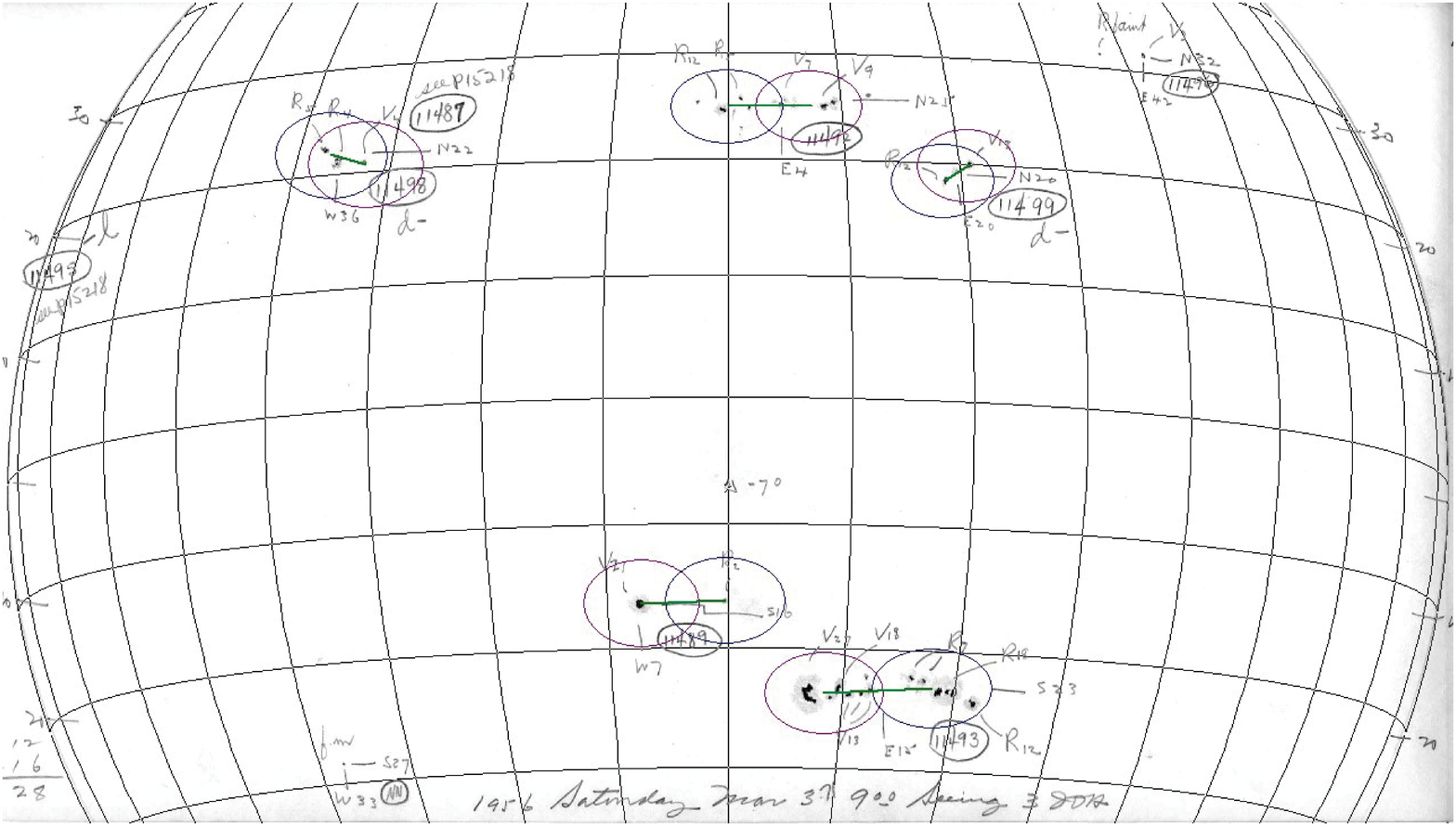}
\caption{Example of bipole pairs identified by our algorithm for observations taken on 3 March 1956. 
Letters V/R corresponds to negative/positive polarities, and the numeric value after the letter 
corresponds to measured field strength in units of hundred Gauss. For example, V27 corresponds to 
negative field strength of 2700 Gauss. Clusters of positive/negative sunspots are outlined by \textit{purple/blue ovals}. 
\textit{Green line} segments correspond to
magnetic axis of each group as determined by our routine. Image is oriented with North up and East 
to the right.}
\label{fig1_new}
\end{figure}

Figure \ref{fig1_new} provides examples of clusters of sunspots and corresponding tilts 
determined using our approach. In this Figure the reader may see  examples of active 
regions, whose tilt follows Joy's law (\textit{e.g.}, group 11499) or deviates from  it (group 11498). There are also cases, when sunspots were not
selected by our routine for tilt determination. For example, non-numbered (NN) pore at
S27W33 does not have polarity information. Group 11496 at N32E42 has its leading polarity 
pore marked as ``R faint'', but this pore was not included in the digitized set leaving the 
negative (V) polarity pores without a bipolar pair.

Each drawing was treated independently of all others. Thus, for example, if an 
active region 
persisted for several days, the dataset will include multiple tilt measurements for 
this region.
This approach may bias long-lived groups. Furthermore, the tilt of an active region 
may change with time
depending on total magnetic flux/area of bipoles (\opencite{McClintock.Norton2016}).
Not all features 
identified on the drawings will contribute to ``bipole pairs'' (for 
comparison, our data set contains about 5$\times$10$^5$ individual flux 
elements, but only about 5$\times$10$^4$ bipoles). Nevertheless, we expect 
that the average orientation of individual pairs belonging to the same group 
would be close to an average orientation of the group.
Using sunspot drawings for tilt studies has some limitations. For example, 
groups with highly asymmetric polarity distributions 
(unipolar sunspots) will be excluded. 
Also, drawings are not inclusive. Some features 
(small pores) may be present on a drawing, but have no corresponding magnetic field measurements. Such 
features will not be included in our determination of tilt.

\begin{figure}[ht]
\includegraphics[width = \textwidth]{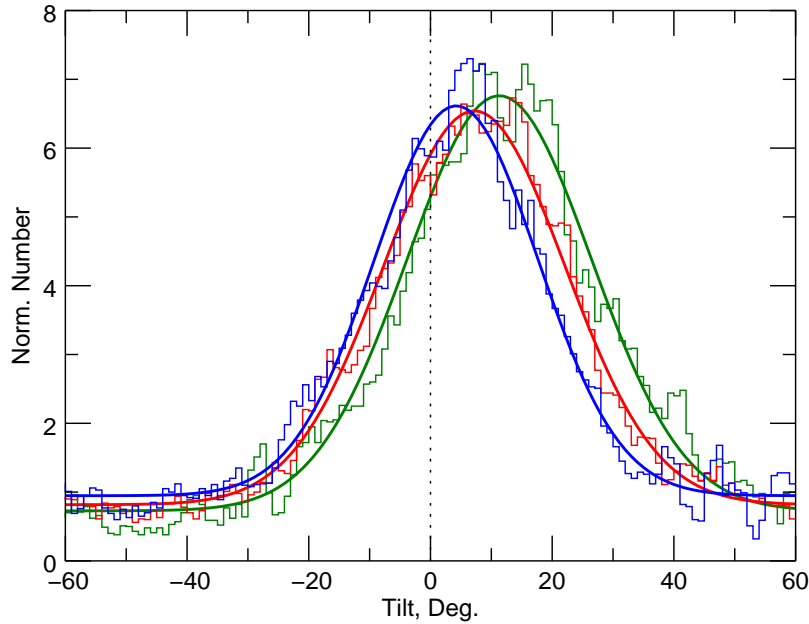}
\caption{Distribution of tilt angles for three latitudinal ranges in the 
Northern hemisphere: 0--10$^\circ$ (\textit{blue}), 10--20$^\circ$ (\textit{red}), and 
20-30$^\circ$ (\textit{green}). \textit{Thin lines} show histograms of tilts at one-degree bins, and \textit{solid 
thick lines} are Gaussian fits to the distributions. For a reference, \textit{vertical dotted 
line} corresponds to zero mean.
To simplify their visual comparison, all three 
distributions are scaled to be about the same maximum amplitude.\label{fig1}}
\end{figure}

\section{Latitudinal Profile of Active Region Tilts}\label{sec:1hemi} 

\begin{figure}[ht!]
\includegraphics[width = \textwidth]{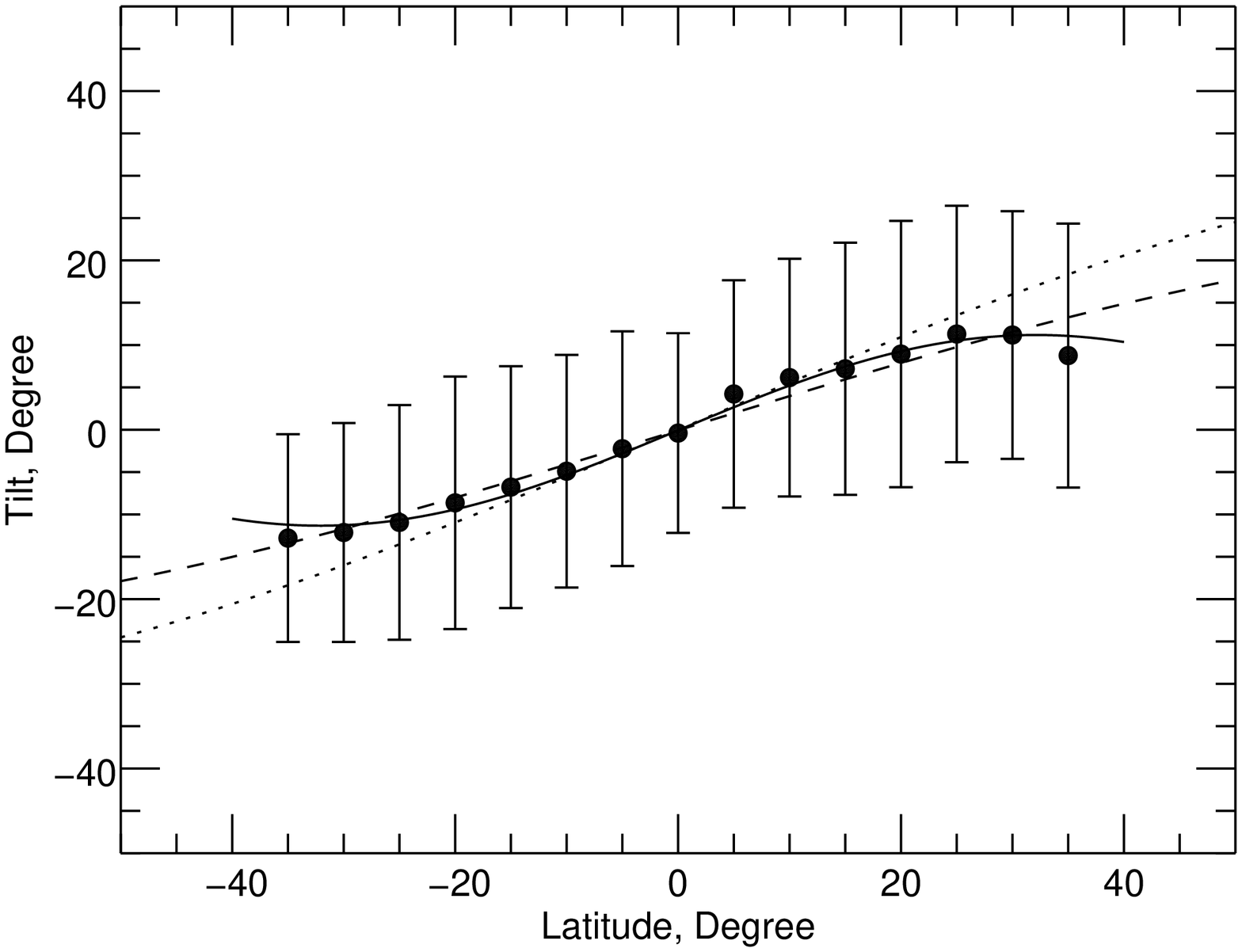}
\caption{Mean tilts of magnetic bipoles (\textit{filled circles}) and their 
standard deviations (error bars) for all cycles. \textit{Solid curve} corresponds to sine
fit by Equation \ref{eq2}, \textit{dashed line} is fit by Equation \ref{eq1}, and \textit{dotted curve} 
represents sine fit by \inlinecite{Stenflo.Kosovichev2012}.}
\label{fig2}
\end{figure}

Next, we divided all identified bipoles in 10-degree latitudinal 
intervals, and used the distributions of tilts in each interval to derive 
the mean tilt and its standard deviation. This approach takes into account any change in tilt 
angle with age of an active region.
The latitudinal intervals were 
spaced by 5 degree in latitudes and thus, neighboring intervals overlap 
by 5 degrees (\textit{e.g.}, 40$^\circ\pm$5$^\circ$, 35$^\circ\pm$5$^\circ$, 
30$^\circ\pm$5$^\circ$ etc). Figure \ref{fig1} provides example of 
distribution of tilts for three latitudinal ranges: 5$^\circ\pm$5$^\circ$, 15$^\circ\pm$5$^
\circ$ and 25$^\circ\pm$5$^\circ$, as well as the corresponding fits of these distributions
by the Gaussian functions. Mean tilts and 
their standard deviations derived that way are listed in Table \ref{tab1}.
Figure \ref{fig2} shows the tilts for all years (see columns 2 and 3 of Table 
\ref{tab1}) in graphical form.

\begin{table}
\caption{Parameters of Gaussian distribution of bipole tilts for selected latitudinal ranges and separately for odd and even cycles}
\label{tab1}
\begin{tabular}{r|rcrcrrcr}
\hline\noalign{\smallskip}
Latitude&\multicolumn{2}{c}{All Cycles}&\multicolumn{3}{c}{Odd Cycles}&\multicolumn{3}{c}{Even Cycles}\\
degree & $\gamma^{a}$ & $\sigma^{b}$ & 
$\gamma^{a}$ & 
$\sigma^{b}$ & 
N$^{c}$ & $\gamma^{a}$ & 
$\sigma^{b}$ & N$^{c}$\\
\hline\noalign{\smallskip}
35&8.77&15.59&7.20&22.70&329&12.30&30.30&413\\
30&11.19&	14.62&	10.60&	26.40&	1142&	13.40&	24.50&	1251\\
25&	11.31&	15.14&	11.40&	24.40&	2699&	12.10&	24.10&	2780\\
20&	8.94&	15.72&	9.70&	24.80&	4915&	9.90&	24.00&	4603\\
15&	7.20&	14.89&	7.90&	23.40&	6374&	7.60&	23.20&	5846\\
10&	6.16&	14.03&	7.10&	21.50&	5363&	5.40&	21.20&	5128\\
5&	4.23&	13.43&	6.20&	21.70&	2984&	3.10&	21.20&	2643\\
0&	-0.39&	11.80&	5.50&	24.00&	1465&	-3.60&	23.50&	1373\\
-5&	-2.23&	13.85&	-0.30&	22.60&	2840&	-5.10&	20.30&	3088\\
-10&	-4.89&	13.73&	-4.50&	22.50&	5406&	-5.80&	21.20&	5668\\
-15&	-6.77&	14.28&	-6.80&	22.30&	6045&	-7.40&	23.40&	6069\\
-20&	-8.63&	14.91&	-8.80&	22.30&	4903&	-9.60&	26.80&	4178\\
-25&	-10.94&	13.86&	-11.70&	22.20&	3100&	-11.00&	26.00&	1991\\
-30&	-12.13&	12.93&	-13.30&	22.10&	1429&	-11.10&	23.00&	812\\
-35&	-12.79&	12.27&	-14.00&	23.10&	384&	-10.10&	21.40&	295\\
\hline\noalign{\smallskip}
\end{tabular}
\\{(a) Average tilt angle}\\
{(b) Standard deviation}\\
{(c) Number of data points}
\end{table}

Similar to previous studies (for review, see \opencite{Pevtsov.etal2014})
the tilt angles of bipoles in our dataset exhibit a slight dependence on latitude 
in their tilt angles 
with a significant scatter. Best linear fit 
to our data (using standard deviations as statistical uncertainties, returns 
\begin{equation}
\gamma = (0.41\pm 0.18) \varphi + (0.00\pm 0.06),
\label{eq1}
\end{equation}
where the coefficients are expressed in radians 
for better comparison with the summary Table 1 in \inlinecite{Pevtsov.etal2014}. 
This fit is shown in Figure \ref{fig2} by dashed line. The coefficients 
(slopes) are similar (albeit slightly larger) to 
\inlinecite{Fisher.etal1995,Dasi-Espuig.etal2010,Ivanov2012}, and the fitted slope is 
significantly 
smaller in comparison with \inlinecite{Stenflo.Kosovichev2012}, (for review, 
see summary Table 1 in \opencite{Pevtsov.etal2014}). 
\inlinecite{Stenflo.Kosovichev2012} fit (shown as a dotted curve in Figure \ref{fig2})
appears to match our data quite well at low latitudes, but at latitudes 
higher that 20 degrees the fitted 
curve  deviates significantly from the observations. Finally, we found 
that the data shown in Figure \ref{fig2} are best represented by
\begin{equation}
\gamma = (0.20\pm 0.08) \sin (2.80 \varphi) + (-0.00\pm 0.06),
\label{eq2}
\end{equation}
where $\gamma$ and $\varphi$ are in radians.
This fit is shown in Figure \ref{fig2} by a solid curve. Here the 2.80 coefficient 
is also 
determined, together with the other two coefficients, via multidimensional minimization, 
using the downhill simplex method (``AMOEBA'' routine, \opencite{Press.etal1992}).

\section{Latitudinal variation of tilt for odd and even cycles}\label{sec:2hemi}

\begin{figure}[ht!]
\includegraphics[width = \textwidth]{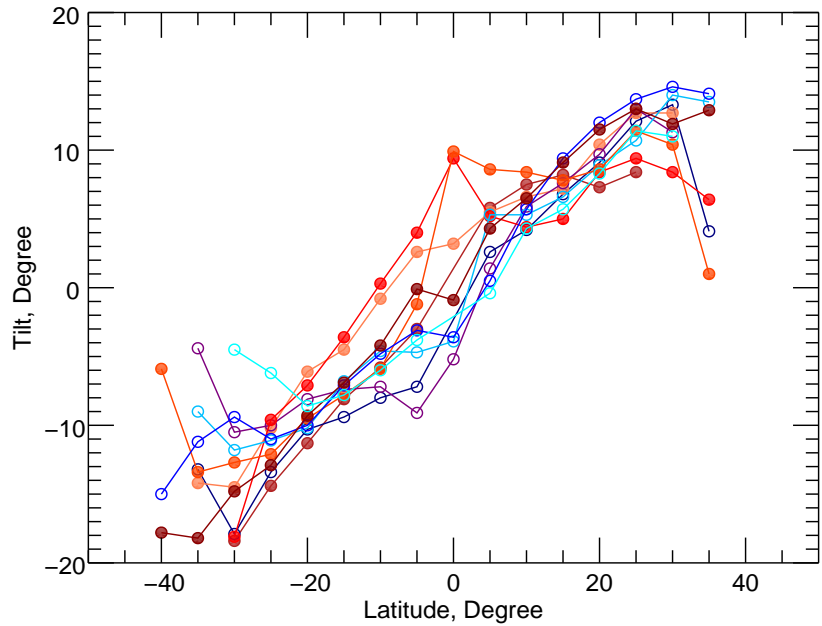}
\caption{Latitudinal profiles of tilts for odd (\textit{filled circles, red-hued color}) and even (\textit{open circles, blue-hued colors}). The data come from Table \ref{tab2}.}
\label{fig3}
\end{figure}

Our dataset allows investigating possible variations in Joy's law with the 
solar cycle. The latitude, at which the solar cycle starts, correlates 
with its cycle amplitude (see, \textit{e.g.} \opencite{Tlatov.Pevtsov2010}): 
high amplitude cycles have
sunspots appearing first at higher latitudes as compared to cycles with 
lower amplitude. Normalizing tilt angles by latitude would then reduce 
this dependence (\textit{e.g.}, $\frac{\gamma}{\varphi} = \frac{A \sin \varphi} 
{\varphi} + \frac{B}{\varphi} \approx \frac{A \varphi} {\varphi} = A.$ 
Thus, coefficient {\rm A} (slope of tilt \textit{vs.} latitude dependence) 
can be used to investigate the observational evidence of 
dependence of tilt on strength of solar cycle. Figure \ref{fig3} shows 
latitudinal tilt profiles for
individual cycles. 
Within the latitude range of $\pm$ 10-25$^\circ$, the tilt profiles appear to be
about the same for Cycles 15--24. At high 
latitudes, the data show significant scatter and large variations 
between different cycles. Coefficients for Equation \ref{eq1} fitted
to individual cycles confirm this visual impression (see, columns 2 and 3 
Table \ref{tab3}; 
To mitigate the differences in tilts at high- and low-latitudes, 
for this test, the fitting was limited to tilts at $\pm$ 10-25$^\circ$ 
latitudinal ranges).
Indeed, the tilt dependence on latitude does differ for different
cycles. For example, on average, the latitudinal dependence is slightly  
steeper for Cycles 15, 16, 18, 20, 21-23 (group 1) 
as compared to Cycles 17, 19, 
and 24 (group 2). However, within each group the difference in A (slope) 
is about the same, and the difference between 
two groups is not too large. Furthermore, the data show no clear 
pattern implying the presence of a relation between
larger A of Cycle $\it n$ and the 
strength of Solar Cycle $\it (n+1)$. For example, Cycles 22 and 23 have 
the steepest latitudinal dependence of all cycles, but neither was 
followed by a particular strong (in amplitude) sunspot cycle. Spearman's 
rank correlation $\tau$ between $\rm A$-coefficient in Cycle ({\it n}) and sunspot 
number 
(SSN) in  Cycle ({\it n+1}) is even negative, $\tau$=$-$0.409 
with significant chance of random occurrence 0.25.

\begin{figure}[ht!]
\includegraphics[width = \textwidth]{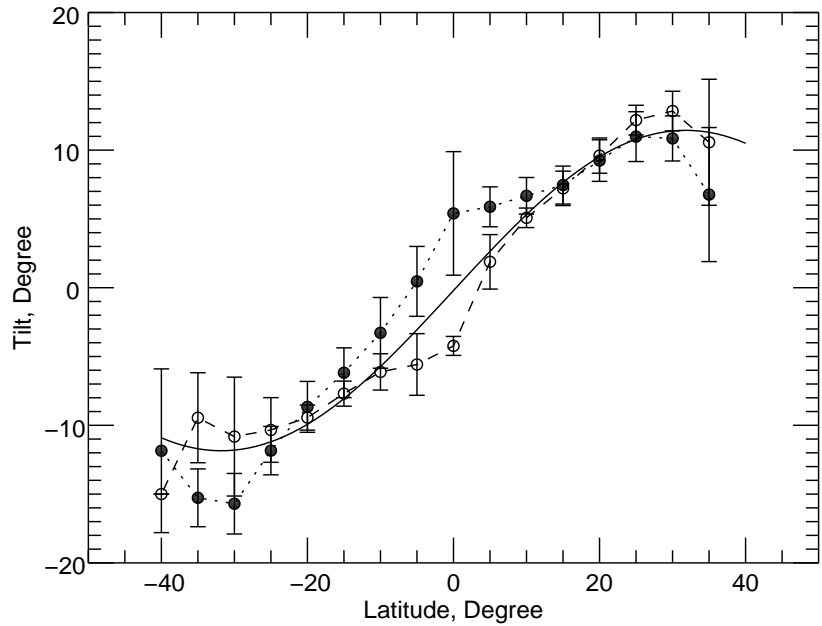}
\caption{Mean tilts of magnetic bipoles for odd (\textit{filled circles, dotted line}) and even (\textit{open circles, dashed line}). \textit{Solid line} represents fit to all dataset shown in Figure \ref{fig1}.}
\label{fig4}
\end{figure}

One interesting result can be noticed in Figure \ref{fig3} at low 
latitudes: odd cycles 
seem to exhibit a positive offset in 
their zero-latitude tilt, while for the even cycles the offset is 
negative. The pattern can also be identified in Table \ref{tab2} in 
tilts at zero latitude. Figure \ref{fig4} shows average latitudinal
profile of tilt angles computed separately for even and odd cycles.
The difference barely exceeds one sigma statistical significance
level, but it does appear non-random.

\begin{table}
\caption{Mean tilts for latitudinal ranges for Cycles 15--24}
\label{tab2}
\begin{tabular}{r|rrrrrrrrrr}
\hline\noalign{\smallskip}
Latitude & C15 & C16 & C17 & C18 & C19 & C20 & C21 & C22 & C23 & C24\\
\hline\noalign{\smallskip}
35&&4.1&&&6.4&13.5&1.0&14.1&12.9\\	 
30&&13.3&12.7&11.3&8.4&14.0&10.40&14.6&11.9&11.0\\
25&8.4&12.1&12.70&13.0&9.4&10.7&11.4&13.7&13.0&11.4\\
20&7.3&9.1&10.4&9.7&8.4&8.9&8.6&12.0&11.5&8.3\\
15&8.2&6.8&7.2&7.6&5.0&6.6&7.80&9.4&9.1&5.7\\
10&7.5&4.2&6.6&5.9&4.4&5.3&8.4&5.7&6.5&4.3\\
5&5.8&2.6&5.5&1.4&5.2&5.3&8.6&0.5&4.3&-0.4\\
0&&&3.2&-5.2&9.4&-3.9&9.9&-3.6&-0.9\\
-5&-3.0&-7.2&2.6&-9.1&4.0&-4.7&-1.2&-3.1&-0.1&-3.8\\
-10&-5.8&-8.0&-0.8&-7.2&0.3&-4.6&-5.9&-4.8&-4.2&-6.0\\
-15&-8.1&-9.4&-4.5&-7.4&-3.6&-6.8&-7.8&-7.1&-6.9&-7.8\\
-20&-11.3&-10.3&-6.1&-8.1&-7.1&-10.2&-9.5&-10.0&-9.3&-8.6\\
-25&-14.4&-13.4&-10.2&-10.0&-9.6&-11.1&-12.1&-11.0&-12.9&-6.2\\
-30&-18.4&-17.9&-14.5&-10.5&-18.1&-11.8&-12.7&-9.4&-14.8&-4.5\\
-35&&-13.2&-14.2&-4.4&&-9.0&-13.4&-11.2&-18.2\\	 
-40&&&&&&&-5.9&-15.0&-17.8&\\	 
\hline\noalign{\smallskip}
\end{tabular}
\end{table}

\begin{table}
\caption{Coefficients of latitudinal fit of Equation \ref{eq1}
(within the latitude range of $\pm$ 10-25$^\circ$)
by solar cycle and maximum annual sunspot number} \label{tab3}
\begin{tabular}{r|rrr}
\hline\noalign{\smallskip}
Cycle & A$\pm \sigma$ & B$\pm \sigma$ & SSN\\
\hline\noalign{\smallskip}
15&   0.50$\pm$0.04&  -0.02$\pm$0.01&175.7\\
16&   0.53$\pm$0.02&  -0.02$\pm$0.01&130.2\\
17&   0.44$\pm$0.02&   0.03$\pm$0.01&198.6\\
18&   0.49$\pm$0.03&   0.01$\pm$0.01&218.7\\
19&   0.36$\pm$0.03&   0.02$\pm$0.01&285.0\\
20&   0.47$\pm$0.01&  -0.00$\pm$0.00&156.6\\
21&   0.50$\pm$0.03&   0.00$\pm$0.01&232.9\\
22&   0.53$\pm$0.01&   0.02$\pm$0.00&212.5\\
23&   0.53$\pm$0.01&   0.01$\pm$0.00&180.3\\
24&   0.41$\pm$0.04&   0.00$\pm$0.01&116.4\\
\hline\noalign{\smallskip}
\end{tabular}
\end{table}

\section{Discussion} \label{sec:discuss}

The orientation (tilt) of active regions is one of critical parameters
that affects the poleward transport of magnetic field originating from decaying active
regions. In  the framework of surface flux transport models, which are 
increasingly used in evaluating the amplitude of future solar cycles,  
active region tilt is one of the important ingredients that affects 
the strength of polar field. In its turn, the strength of polar field 
determines the amplitude of next solar cycle (\textit{e.g.},\opencite{Baumann.etal2004,Cameron.etal2010}).
Thus, a proper determination of latitudinal variation of active region tilt angles,
and its possible change from cycle to cycle are important for understanding and modeling future 
sunspot activity cycles.
Our study revealed several previously unknown 
tendencies in the variation of the tilt angles of solar active regions
with latitude (so called Joy's law): the presence of a maximum at 
about 25--30$^\circ$ range of
latitudes and an opposite offset in non-zero tilt at the equator for odd and 
even cycles.

In some previous studies, one of the unspoken assumptions was that 
the functional dependence of tilts is a function f($\varphi$), 
which monotonically increases with latitude. Some studies used 
f($\varphi$) $\propto \varphi$ or f($\varphi$) $\propto \sin \varphi$. 
By itself, the assumption is reasonable, if one accepts that the action 
of the Coriolis force is the underlying explanation for this solar 
phenomenon. In fact, the combination of the Coriolis force and fluxtube dynamics (\textit{e.g.}, material drainage, significant scatter presumably 
caused  by interaction with turbulent convection) seems to explain the 
previously observed properties of active region tilts reasonable 
well (\opencite{Fisher.etal2000}). 

However, the non-monotonic behavior in tilt angle 
with latitudes described in this paper, may still be explained in the 
framework of \inlinecite{Fisher.etal1995} model. On the basis of a model of
a thin flux tube, whose main axis is distorted by the action of the 
Coriolis force and by the interaction with turbulent convection,  
\inlinecite{Fisher.etal1995} derived a dependence of tilt of active regions 
on latitude and footpoint separation. This dependence may, in fact, 
offer at least a quantitative explanation for our observation of a 
non-monotonic behavior of tilt with latitude. Active regions emerging 
at the  beginning of each solar cycle, are typically smaller in size 
and they emerge at higher latitudes. The main activity in each cycle 
develops in lower latitudes with large regions usually emerging around 
the maximum of cycle or slightly after it. This range of latitudes 
will exhibit the largest tilt. As smaller regions tend to 
exhibit smaller tilt, those regions that emerge at higher latitudes 
at the beginning of the cycle will exhibit tendency for smaller tilt, as compared with the larger regions in 
lower latitudes that emerge later in the 
cycle. The number of active  regions at high latitudes is 
significantly smaller as compared to lower latitudes (see, Table 
\ref{tab1}, columns 6 and 9), which increases the statistical uncertainty of 
tilt angles 
for high latitudes. We think that a combination of these two aspects 
could explain the appearance of a maximum in tilt angles at $\pm$ 
25-30$^\circ$ range of
latitudes.

The reader should note, however, that while we invoke the action of the 
Coriolis force 
to explain a non-monotonic behavior of active region tilts with latitude, 
we do not argue that this effect is the only explanation for the observed 
orientation of active regions. For example, \inlinecite{Kosovichev.Stenflo2008} 
see the relaxation of tilt 
after the active region emergence to some mean value for a given latitude 
(and not strictly to 
East-West direction) as an indication that tilt must represent the 
orientation of toroidal field in the convection zone.
The functional dependence of tilt on sine of latitude fitted by
Equation \ref{eq2} should only be used for the range of latitudes hosting
active regions ($\approx \pm$45$^\circ$). The reader could notice
that at the latitude of about 64--65$^\circ$ the fitted curve with cross 
zero and reverse its sign.

The presence of an offset in the non-zero tilt at solar equator is a clear 
indication that the Coriolis force alone cannot explain the active 
region tilt. The most plausible explanation is the kink instability 
inside the flux tubes that define the active regions. The direction 
of the kink would be determined by the sense of the internal twist 
(helicity) inside the magnetic flux tube. In fact, earlier 
investigations do find that for some active regions, the sign-relation 
between active region tilt and internal twist of their magnetic fields  
supports the interpretation of tilt as the result of kink-instability
(\opencite{Canfield.Pevtsov1998,Holder.etal2004,Tian.etal2005}).
There is also a possibility that the non-zero tilt at solar equator might be a 
consequence of the (sunspot) magnetic equator not being aligned with the solar 
rotation equator. The idea of sunspot-based magnetic equator was introduced by 
\inlinecite{Pulkkinen.etal1999}, who defined it as an average latitude of sunspots of the 
northern and southern hemispheres. They also found the location of this sunspot 
magnetic equator to switch between the northern and southern hemisphere with a 
period of about 90 years. \inlinecite{Zolotova.etal2009} verified this hemispheric 
asymmetry in sunspot activity, with magnetic equator being offset to the south 
during Cycles 12-15, to the north during Cycles 17-19 and back to the south during 
Cycles 20-23. However, since these results find that the sunspot equator oscillates 
at a period of several solar cycles, they cannot explain the present observations. 
Instead, we note of the hemispheric asymmetry of the streamer belt, which was found 
to oscillate in a 22-year cycle (\opencite{Mursula.Zieger2001,Mursula.etal2002}). 
Accordingly, the streamer belt is displaced northward or southward in alternating 
cycles.  The alternating  occurrence of positive (negative) tilts at the equator in 
odd (resp. even) cycles may lead to a corresponding north-south displacement of the 
streamer belt later in the cycle. This possible connection will be elaborated later 
in more detail. Anyway, this is the only north-south asymmetric phenomenon close to 
the solar equator, which alternates systematically from cycle to cycle.

Finally, although our data do show some variations in the steepness of slope of 
the latitudinal variation of tilts between different cycles, we found
no correlation between steepness of latitudinal dependence of tilt 
angles  in Cycle ({\it n}) and the strength of the following sunspot 
Cycle ({\it n+1}). This lack of correlation is not surprising, as the 
strength of sunspot cycle is affected by several other factors, 
not only by the tilt of active regions. For example, prior conditions 
of polar field (its strength) determine how much  
magnetic flux is required for its reversal 
and re-building for the next cycle.
In addition, the strength of polar field may be affected by the emergence of
active regions with non-Hale polarity orientation (\opencite{Yeates.etal2015}). 
Recent modeling indicates that emergence of a single ``rogue'' region in the right 
phase of solar cycle may significantly reduce the strength of polar field, and in the extreme case,
it may even shutdown the solar cycle (\opencite{Nagy.etal2017}). The effect of ``rogue'' active regions
may weaken a more ``deterministic'' effect of tilt of active regions on strength of polar field.

\begin{acks}
Work of individual co-authors was supported by their 
National grants and projects:
the Academy of Finland to the ReSoLVE Centre of 
Excellence (project no. 272157),
the Russian Foundation for Basic Research (RFBR, project 
18-02-00098), the Russian Science Foundation 
(RSF, project 15-12-20001), and NASA's grant NNX15AE95G.
The authors are members of 
international team on Reconstructing Solar and 
Heliospheric Magnetic Field Evolution Over the Past 
Century supported by the International Space Science 
Institute (ISSI), Bern, Switzerland. The National Solar 
Observatory (NSO) is operated by the Association of 
Universities for Research in Astronomy (AURA), Inc., 
under cooperative agreement with the National Science 
Foundation.
\end{acks}

\smallskip\noindent
{\bf Disclosure of Potential Conflicts of Interest} The authors declare that they have no conflicts of interest.

\bibliographystyle{spr-mp-sola}
\bibliography{tlatova}

\end{article}
\end{document}